\documentclass[twocolumn,showpacs,preprintnumbers,amsmath,amssymb]{revtex4}

\usepackage{graphicx}
\usepackage{dcolumn}
\usepackage{bm}

\def\Journal#1#2#3#4{{#1} {\bf#2}, #3 (#4)}

\def\PRL{Phys. Rev. Lett.}
\def\PRC{Phys. Rev. C}

\def\JPG{J. Phys. G}

\newcommand{\be}{\begin{equation}}
\newcommand{\ee}{\end{equation}}
\newcommand{\bea}{\begin{eqnarray}}
\newcommand{\eea}{\end{eqnarray}}

\begin{document}

\title{Pentaquark Search in Relativistic Heavy Ion Collisions with STAR}
\author{Sevil Salur}
\email{sevil.salur@yale.edu}
\affiliation{%
(for the STAR Collaboration)\\
\\
Yale University\\
Physics Department, 272 Whitney Ave.\\
New Haven, CT 06520\\
}%
\date{\today}
\begin{abstract}
We report on the progress of the pentaquark searches by the STAR
collaboration in p+p, d+Au, and Au+Au at $s_{NN}=\sqrt{200}$ GeV
collisions through one of the decay modes of
$\Theta^{+}\rightarrow p+K_{s}^{0}$. $\Theta^{+}$ state is an
exotic baryon with strangeness $S=1$ and is the lightest isospin
member of the expected antidecuplet. We compare our techniques for
the pentaquark search with those for the short-lived resonances.
These results were presented as a poster at Quark Matter 2004.
\end{abstract}
\maketitle
\section{Introduction}
The first observations of the $\Theta^{+}$ pentaquark, a five
quark bound system of uudd$\overline{s}$, have been reported in
photon-nucleus and kaon-nucleus reactions
\cite{{leps},{clas},{diana}}. The presence of this state was
predicted by R. Jaffe with multiquark bag models
\cite{{jaffe1},{jaffe2}} and later by D. Diakonov et al. using
chiral soliton models of baryons \cite{diakonov}.

The high energies and particle densities resulting from collisions
at the Relativistic Heavy Ion Collider (RHIC) are expected to
favor pentaquark production. The data that we analyze were taken
by the STAR (Solenoidal Tracker At RHIC) experiment, one of the
four relativistic heavy ion experiments at RHIC. The large
acceptance of STAR's Time Projection Chamber (TPC) is ideal for
such rare particle searches. The short lifetimes predicted for
pentaquarks require that a mixing technique be used to reconstruct
pentaquarks via their decay products. This technique has already
been used successfully by STAR to reconstruct and study resonances
\cite{{kstar},{rho}}.
\section{Analysis and Particle Identification}
Charged daughter particles are identified by the momentum and the
energy lost per unit length, $dE/dx$, measured with the Time
Projection Chamber of STAR. For long lived particles ($c\tau\sim$
few cm) such as $K_{s}^{0}$, $\Lambda$ and $\Xi$, the decay vertex
topology information is used for their identification. This method
cannot be used for pentaquarks since they decay strongly
($c\tau_{\Theta}\sim fm$). An alternative method, a mixing
technique, is used successfully to identify short lived particles
such as resonances ($c\tau_{\Sigma(1385)}=5 \;fm$) and can be used
to search for pentaquarks and dibaryons.

In the mixing technique, for example
$\Sigma(1385)\rightarrow\Lambda+\pi$, we identify first the
$\Lambda$'s by their decay vertex topology, due to their long
lifetime ($c\tau_{\Lambda}= 7.89 \;cm$). This $\Lambda$ candidate
is then combined with a $\pi$ to get $\Sigma(1385)$. The
background is described by combining $\pi$'s from one event with
the $\Lambda$'s from another event. Similarly for pentaquarks such
as $\Theta^{+}\rightarrow K_{s}^{0}+p$, we first identify the
$K_{s}^{0}$ via its decay vertex topology and then combine this
$K_{s}^{0}$ candidate with a proton candidate in the same event to
extract the signal and in a different event to describe the mixing
background.
\section{Status of Current Studies}
\subsection{Monte Carlo Studies}
\begin{figure}[h!]
  \includegraphics[width=8cm,height=4cm]{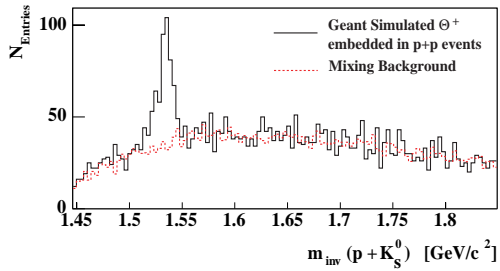}\\
  \caption{Invariant mass spectrum of the Monte Carlo simulated $\Theta^{+}$ embedded in real p+p events
  with the mixing technique. Black solid line is the signal and red dashed line is the mixed event background.
  The simulated signal can be clearly seen.}
  \label{invmassnugget}
\end{figure}

\begin{figure}[h!]
  \includegraphics[width=7.5cm,height=3.5cm]{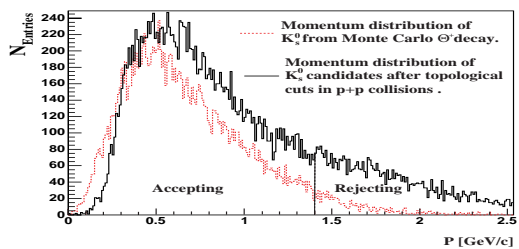}\\
  \caption{$K_{s}^{0}$ momentum distribution: Black solid line is of the accepted
  $K^{0}_{s}$ after the topological cuts and red dashed line is of the decay daughters
   of the Monte Carlo simulated $\Theta ^{+}$ for the same number of events.}\label{fig:K0mom}
\end{figure}

\begin{figure}[h!]
  \includegraphics[width=7.5cm,height=3.5cm]{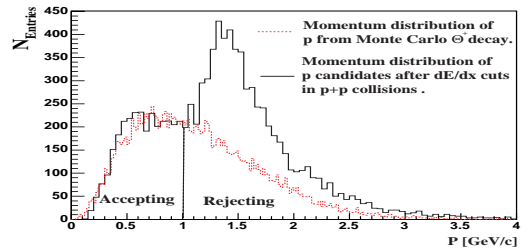}\\
  \caption{Proton momentum distribution: Black solid line is of the accepted
  p after the dE/dx cut and red dashed line is of the decay daughters
  of the Monte Carlo simulated $\Theta ^{+}$ for the same number of events.}\label{fig:Protonmom}
\end{figure}

To study the decay mechanism and optimize the applied cuts, we use
Monte Carlo simulations.   In this study, one Monte Carlo
$\Theta^{+}$ pentaquark is chosen from a thermal exponential
distribution with $T_{inv slope} = 250$ MeV for a rapidity $\mid
y\mid <1.5$ and is embedded in a single real p+p event after a
full TPC simulation. The chosen input width, $10\;MeV/c^{2}$, and
the input mass, $1.54 \;GeV/c^{2}$,
\begin{figure}[h!]
 \includegraphics[width=8cm,height=4cm]{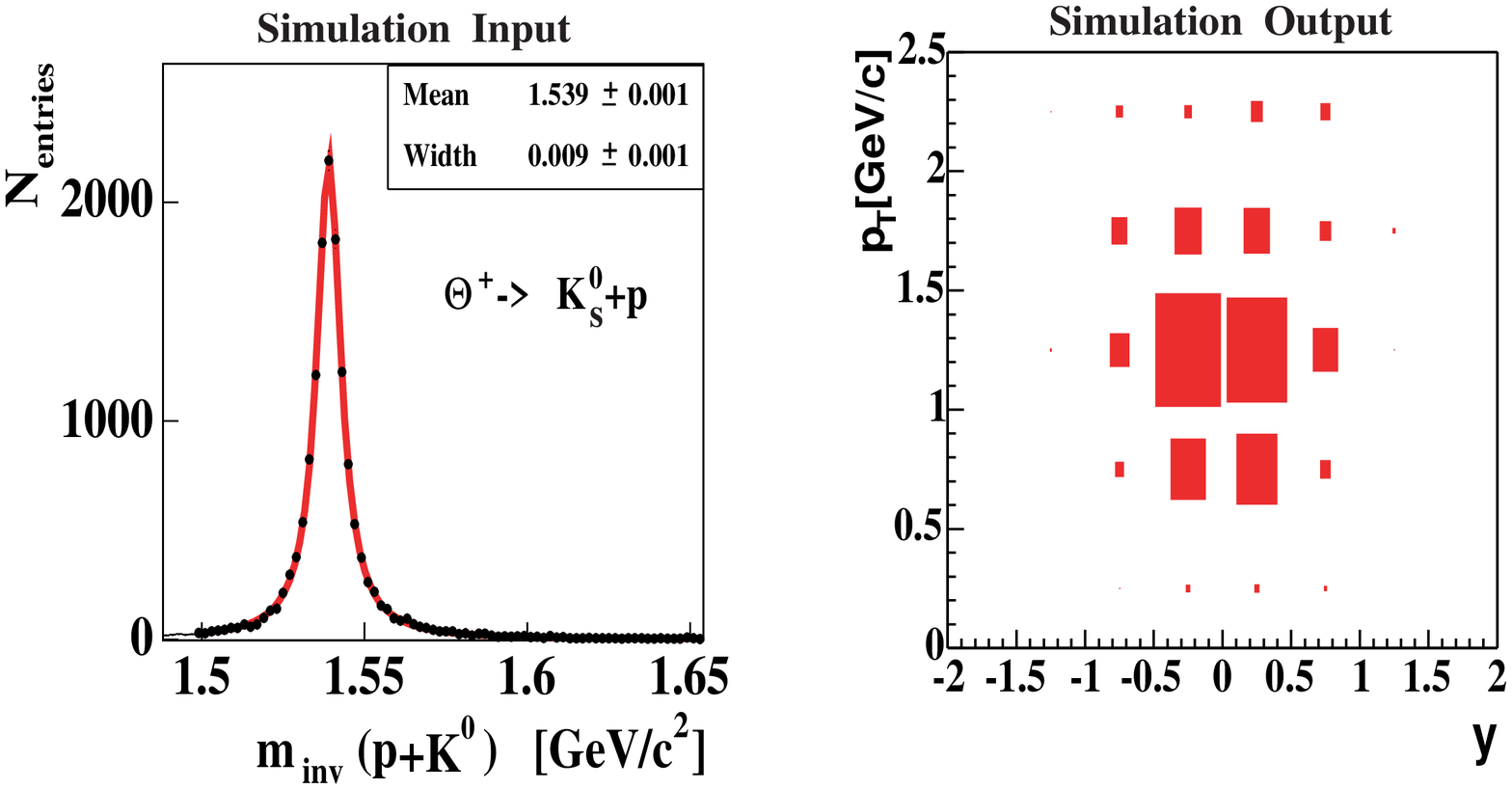}\\
  \caption{Invariant mass spectrum of the simulated Monte Carlo input for the $\Theta^{+}$
  and $p_{T}$ vs rapidity of the $\Theta^{+}$ that can be identified with the TPC.}
\label{fig:simulation}
\end{figure}
are consistent with the observed mass and width of $\Theta^{+}$
\cite{{leps},{clas},{diana}}. In Fig.~\ref{fig:simulation} the
invariant mass spectrum of simulated $\Theta^{+}$ and the TPC
acceptance for the corresponding simulated particles can be found.
In Fig.~\ref{invmassnugget} the invariant mass spectrum of the
reconstructed $\Theta^{+}$ are presented. We find that $\sim3\%$
of these Monte Carlo $\Theta^{+}$'s are successfully reconstructed
with this technique. The reconstructed width and the mass is
consistent with the Monte Carlo input.
We can study the decay properties with the simulated tracks such
as the momentum distribution of the decay daughters. Fig.
\ref{fig:K0mom} and Fig. \ref{fig:Protonmom} are the momentum
distributions of the $K_{s}^{0}$ and p respectively. With this
study we can apply momentum cuts to increase the
signal-to-background ratio. This ratio can be increased clearly by
selecting protons with momentum less than 1~GeV/c. Further
detailed studies on other variables are needed to optimize the
signal-to-background ratio.
\subsection{Mixing Technique Works} 
\begin{figure}[h!]
  \centering
  \includegraphics[width=8cm,height=4cm]{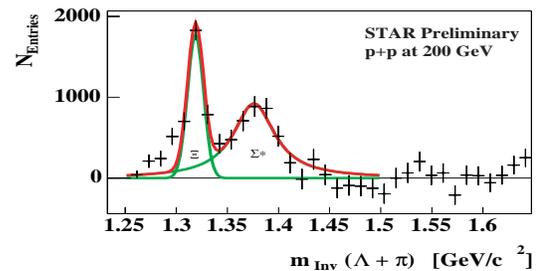}\\
  \caption{Background subtracted invariant mass spectrum of $\Sigma(1385)$ in p+p
  collisions.}\label{fig:ppsigma}
\end{figure}

\begin{figure}[h!]
  \centering
  \includegraphics[width=8cm,height=4cm]{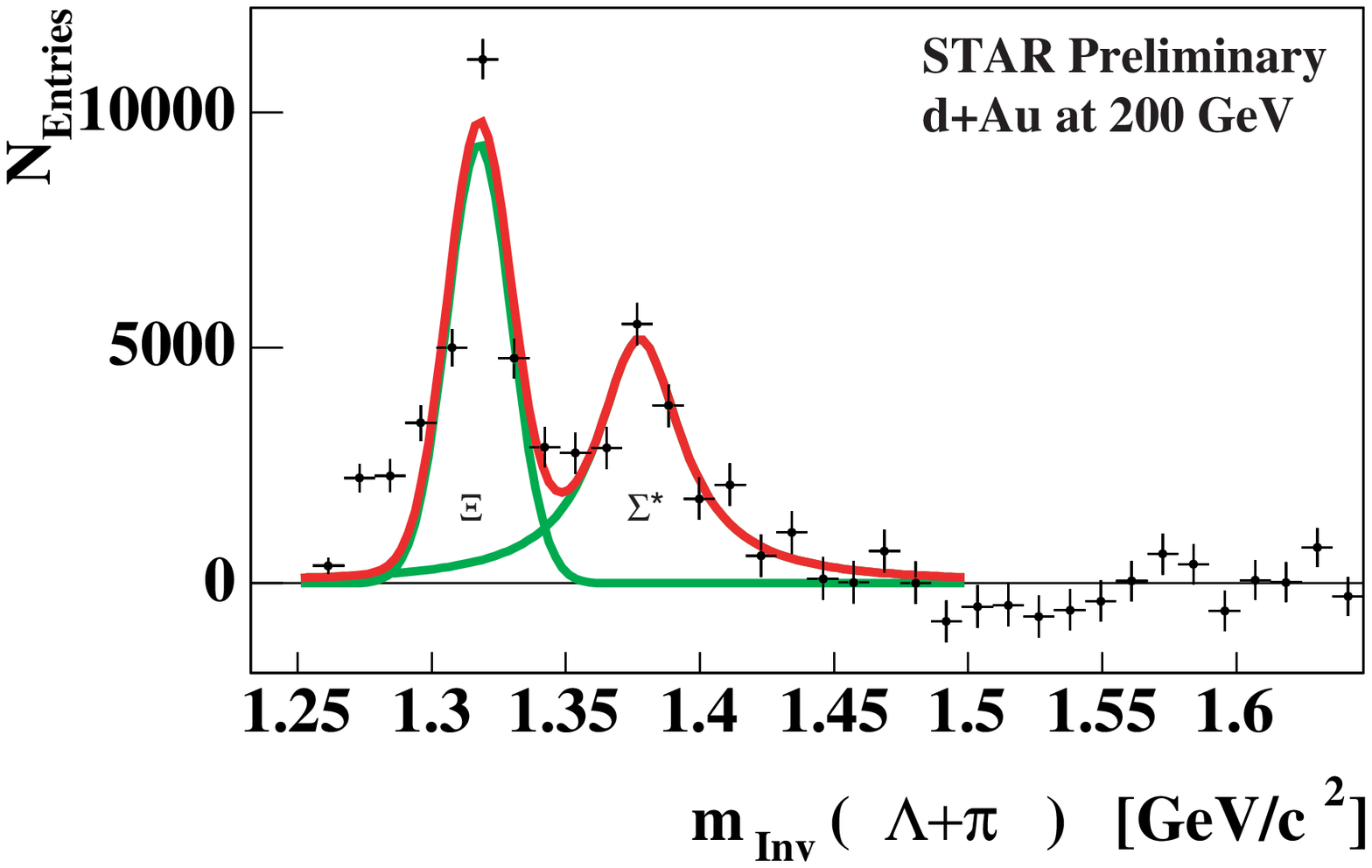}\\
  \caption{Background subtracted invariant mass spectrum of $\Sigma(1385)$ in d+Au
  collisions.}\label{fig:dAusigma}
\end{figure}

\begin{figure}[h!]
  \centering
  \includegraphics[width=8cm,height=4cm]{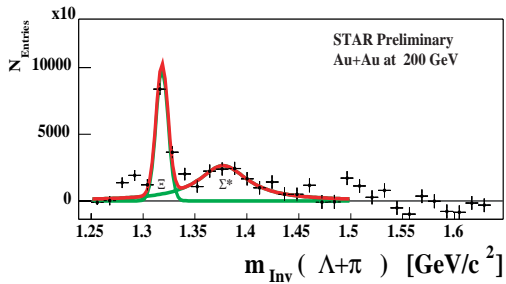}\\
  \caption{Background subtracted invariant mass spectrum of $\Sigma(1385)$ in Au+Au
   central collisions.}\label{fig:AuAusigma}
\end{figure}

$\Sigma(1385)\rightarrow
\Lambda+\pi_{bachelor}$ is identified in p+p, d+Au, and Au+Au
collisions with the mixing techniques. Fig~\ref{fig:ppsigma} and
Fig~\ref{fig:dAusigma} are the preliminary, background subtracted
invariant mass spectra of the $\Xi\rightarrow\Lambda+\pi$ and
$\Sigma(1385)\rightarrow\Lambda+\pi$ for p+p and d+Au collisions
at $\sqrt{s_{NN}}=200$ GeV. In these figures the invariant mass
spectra of $\Sigma(1385)^{+}$, $\Sigma(1385)^{-}$ and their
antiparticles are summed to improve the statistics. Similarly
Fig.~\ref{fig:AuAusigma} is for the Au+Au central collisions. For
this figure only $\Sigma(1385)^{+}$ and $\Sigma(1385)^{-}$ are
added.

The $\Xi$ peaks are fit with gaussian distributions and the
$\Sigma$(1385) peaks are fit with a Breit-Wigner distribution.
Antiparticle-to-particle ratios of $\Xi$ and $\Sigma$(1385) are
observed as $0.89\pm0.04$ and $0.90\pm0.07$ respectively in p+p
collisions. (In addition to the given statistical errors, these
ratios  also have a 20\% systematic error due to the mixed event
background normalization.) These values are consistent with what
we observe for other hadronic particle ratios at RHIC in p+p
collisions, such as the antiproton to proton ratio which is
$\overline{p}/p\sim 0.8$. We expect a similar production of
pentaquarks and their antiparticles at RHIC.
\begin{figure}[h!]
  \centering
  \includegraphics[width=9cm]{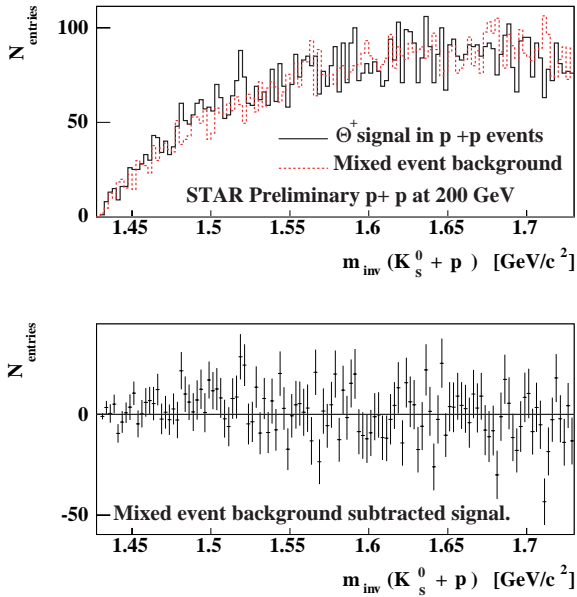}\\
  \caption{Invariant mass spectrum of the $\Theta^{+}$ in p+p
  collisions. Black solid line is the signal and red dashed line is the mixing background
  for the upper figure. The lower figure is the mixed event background subtracted invariant
  mass spectrum. The expected position of the mass of $\Theta$ is at
  $1.54\; GeV/c^{2}$.}\label{fig:nugget}
\end{figure}
A very preliminary invariant mass spectrum of the $\Theta^{+}$ in
p+p collisions is shown Figure \ref{fig:nugget}. Further studies
are needed to optimize the signal-to-background as no strong
signal is observed. The invariant mass spectrum that we observe
for the $\Theta^{+}$ in p+p collision events is consistent with
our predictions for the significance of the signal given our
current statistics. Details can be found in \S
\ref{sect:feasibility}. The signal-to-background depends highly on
the selection of events and applied cuts. To improve cuts and
understand the decay mechanism we can do more detailed simulation
studies. For Au+Au and d+Au collisions similar invariant mass
spectra are observed.
\subsection{Feasibility Studies}\label{sect:feasibility}
Assuming the $\Theta^{+}$ production is $10-100\%$ of the $\Lambda
(1520)$ in p+p collisions, one can predict the yield of
$\Theta^{+}$. Preliminary $dN/dy$ of $\Lambda(1520)$ at
mid-rapidity is 0.004 per event in p+p collisions
\cite{{ludo},{markert}}. There are 8 Million p+p events available
for this analysis. This corresponds to a production of $\sim30000$
$\Lambda(1520)$ and a production range of $\sim3000$ to
$\sim30000$ $\Theta^{+}$'s in these p+p events. As the efficiency
of the mixing technique is $\sim3\%$ and the branching ratio of
the $\Theta\rightarrow K_s^{0}+p$ is $\sim25\%$ (assuming that the
branching ratios of $\Theta\rightarrow K^{0}+p$ and
$\Theta\rightarrow K{+}+N$ are $50\%$ each), 20-200 of the
$\Theta$'s can be found with the mixing technique. The background
pairs per event in the 1.54$\pm 5$ MeV mass range is 3200. This
corresponds to a significance of 0.25 to 3 for the significance
defined as $\frac{Signal}{\sqrt{2\times Background+Signal}}$.
Similarly one can repeat the same study for Au+Au and d+Au
collisions and correspondingly predict a significance of 2-7 for
1.5 Million Au+Au events and 1-16 for 10 Million d+Au for the
predicted production of one $\Theta^{+}$ per rapidity per
collision \cite{{Ko},{Randup},{raf:1}}. To estimate the yield for
the d+Au collisions we assume $N_{part}$ scaling. The mean number
of participants in d+Au is 8, in p+p it is 2, and in Au+Au it is
350 for the most central collisions. The lower limit is obtained
from p+p scaling while the upper limit is from Au+Au yield
estimates.
\section{Conclusion}
Preliminary acceptance and efficiency studies show that we should
be able to find  pentaquarks at the few \% level. Resonances can
be clearly reconstructed via event mixing techniques in p+p, d+Au
and Au+Au central collisions. Optimization of cuts to improve the
signal over background is in progress. There is a possibility of
measuring the anti-pentaquarks at RHIC since the antibaryon to
baryon ratio indicates a nearly net baryon free region
\cite{ratio}. 100 Million events for Au+Au collisions at
$\sqrt{s_{NN}}=200$ GeV are planned to be collected in 2004, Run
4, which has just started. This would be 70 times the currently
available data. The predicted significance of identifying
$\Theta^{+}$ is at a range of 20 to 84 with the forthcoming data.
We expect to put a more definite upper limit to the yields and
production mechanisms of the pentaquarks in Au+Au collisions with
the 2004 run which will finish collecting data at the end of May
2004.


\end{document}